\documentclass[pra, aps, showpacs, floatfix, notitlepage, reprint, nofootinbib,longbibliography]{revtex4-2}

\usepackage{amsmath}
\usepackage{graphicx}
\usepackage[ansinew]{inputenc}
\usepackage{array}
\usepackage{color}
\usepackage[caption=false,justification=justified]{subfig}

\usepackage[normalem]{ulem}

\usepackage{amsxtra}
\usepackage{amstext}
\usepackage{amssymb}
\usepackage{latexsym}
\usepackage{dsfont}
\usepackage{lipsum}
\usepackage[colorlinks=true,allcolors=blue]{hyperref}

\usepackage{enumitem}
\setenumerate[1]{label=(\Roman*)}

\newcommand{\mb}[1]{\mathbf{#1}}

\newcommand{\tm}[1]{\text{#1}}
\newcommand{\lel}{\left}
\newcommand{\rer}{\right}

\newcommand{\bra}[1]{\left\langle{#1}\right|}
\newcommand{\ket}[1]{\left|{#1}\right\rangle}
\newcommand{\comm}[2]{\left[#1,#2\right]}

\newcommand{\ah}{\hat{a}}
\newcommand{\adh}{\hat{a}^\dagger}
\newcommand{\bh}{\hat{b}}
\newcommand{\bdh}{\hat{b}^\dagger}
\newcommand{\ch}{\hat{c}}
\newcommand{\cdh}{\hat{c}^\dagger}
\newcommand{\ph}{\hat{p}}
\newcommand{\pdh}{\hat{p}^\dagger}

\newcommand{\bk}{\mathbf{k}}
\newcommand{\cL}{\mathcal{L}}
\newcommand{\br}{\mathbf{r}}

\newcommand{\bx}{\mathbf{x}}
\newcommand{\cdd}{\cdot}
\newcommand{\la}{\lambda}
\newcommand{\om}{\omega}
\newcommand{\omt}{\widetilde{\omega}}

\newcommand{\thh}{\theta}

\newcommand{\al}{\alpha}
\newcommand{\altt}{\widetilde{\al}}
\newcommand{\be}{\beta}
\newcommand{\bett}{\widetilde{\be}}
\newcommand{\ga}{\gamma}
\newcommand{\gatt}{\widetilde{\ga}}

\newcommand{\dive}[1]{\boldsymbol{\nabla}_{#1}\cdot}
\newcommand{\grad}[1]{\boldsymbol{\nabla}_{#1}}

\def\XXint#1#2#3{{\setbox0=\hbox{$#1{#2#3}{\int}$}
     \vcenter{\hbox{$#2#3$}}\kern-.5\wd0}}

\newcommand{\intk}{\int d^3k \;}
\newcommand{\intkf}{\int \frac{d^3k}{(2\pi)^{3/2}} \;}

\newcommand{\intkp}{\int d^3k' \;}

\newcommand{\intr}{\int d^3r \;}

\newcommand{\Ah}{\mathbf{A}}
\newcommand{\Ch}{\mathbf{C}}
\newcommand{\Eh}{\mathbf{E}}
\newcommand{\Dh}{\mathbf{D}}
\newcommand{\Hh}{\mathbf{H}}
\newcommand{\Bh}{\mathbf{B}}

\newcommand{\Ph}{\mathbf{P}}
\newcommand{\Mh}{\mathbf{M}}

\newcommand{\Pih}{\boldsymbol{\Pi}}
\newcommand{\Ahh}{\mathbf{\widehat{A}}}
\newcommand{\Chh}{\mathbf{\widehat{C}}}
\newcommand{\Ehh}{\mathbf{\widehat{E}}}
\newcommand{\Dhh}{\mathbf{\widehat{D}}}
\newcommand{\Hhh}{\mathbf{\widehat{H}}}
\newcommand{\Bhh}{\mathbf{\widehat{B}}}

\newcommand{\Phh}{\mathbf{\widehat{P}}}
\newcommand{\Mhh}{\mathbf{\widehat{M}}}

\newcommand{\Pihh}{\boldsymbol{\widehat{\Pi}}}

\begin{document}
\title{Quantised helicity in optical media}
\author{Neel Mackinnon}
\author{J{\"o}rg B. G{\"o}tte}
\author{Stephen M. Barnett}
\author{Niclas Westerberg}
\email{Niclas.Westerberg@glasgow.ac.uk}
\affiliation{School of Physics and Astronomy, University of Glasgow, Glasgow, G12 8QQ, United Kingdom}
\begin{abstract}
We present a new approach to the definition of optical helicity in a medium. Our approach resolves the problem that duality transformations which simultaneously combine $\mathbf{E}$ with $\mathbf{H}$ and $\mathbf{D}$ with $\mathbf{B}$ are incompatible with linear constitutive relations. We find that the helicity density in a medium, as the conserved quantity associated with duality transforms, must contain an explicit contribution associated with the polarisation and magnetisation of the matter, and that it can be expressed naturally in terms of the elementary polarised excitations of the system. In media for which the helicity is conserved, each circular excitation carries a well-defined helicity. However, in a medium for which the helicity is not conserved, we find that the time-varying helicity can be viewed  in terms of oscillations between different helicity eigenstates, analogous to neutrino oscillations. Here we explicitly study the helicity in homogeneous and lossless media but we believe that, differently to other choices, this helicity is readily generalisable to media that may be inhomogeneous, lossy, chiral or nonreciprocal.
\end{abstract}

\date{\today}
\maketitle

\section{Introduction}\label{sec:intro}

\noindent Helicity is an important and well-understood property in particle physics \cite{Weinberg, Srednicki}, and in optics, where it is associated with circular polarisation \cite{BornAndWolf}. The helicity is the fundamental measure of the handedness of light, and as such it has been identified as the key optical quantity in the differential action of light on chiral molecules and media \cite{molecularQEDbook, barron, chiral2}. This includes, in particular, chirally-selective forces present in a field with a helicity gradient \cite{Cameron2014,Brasselet2019}.
 
At low energies, that is in the absence of free charges and currents, Maxwell's equations reduce to
\begin{align}
\boldsymbol\nabla\cdot\mathbf{D}&=0,\nonumber &
\boldsymbol\nabla\cdot\mathbf{B}&=0,\nonumber\\
\boldsymbol\nabla\times\mathbf{E}&=-\partial_t\mathbf{B}, &
\boldsymbol\nabla\times\mathbf{H}&=\partial_t\mathbf{D}.\label{eq:MaxwellsFreeSpace}
\end{align}
It was Heaviside and Larmor \cite{HeavisideDuality,LarmorDuality} who first pointed out that these equations possess an electric-magnetic symmetry in that they are unchanged if we perform the rotations
\begin{align}\label{eq:dualityTransformDB}
\mathbf{E}'= \cos{\theta}\, \mathbf{E}+\sin{\theta}\,\mathbf{H},\nonumber\\
\mathbf{H}'= \cos{\theta}\, \mathbf{H}-\sin{\theta}\,\mathbf{E},\nonumber\\
\mathbf{D}'= \cos{\theta}\, \mathbf{D}+\sin{\theta}\,\mathbf{B},\nonumber\\
\mathbf{B}'= \cos{\theta}\,\mathbf{B}-\sin{\theta}\, \mathbf{D},
\end{align}
where $\theta$ is a pseudoscalar, and we work in natural units in which $\varepsilon_0$, $\mu_0$, $c$ and $\hbar$ are all unity. In free space we can associate this symmetry with a locally conserved helicity density \cite{Nienhuis2016}
\begin{equation} \label{eq:helicityFreeSpace}
h_\tm{free}=(\mathbf{A}\cdot\mathbf{B}-\mathbf{C}\cdot\mathbf{D})/2,
\end{equation}
where $\mathbf{A}$ is the magnetic vector potential, defined by $\mathbf{B}=\boldsymbol{\nabla}\times\mathbf{A}$, and $\mathbf{C}$ is the electric vector potential, defined by $\mathbf{D}=-\boldsymbol{\nabla}\times\mathbf{C}$ \cite{bateman, stratton}. The flux or flow of the helicity is simply the spin density \cite{Barnett2012,Nienhuis2016}. Indeed, in free space the Heaviside-Larmor transformation \eqref{eq:dualityTransformDB} may be interpreted as a rotation of the fields for each plane wave through an angle $\theta$ about its direction of propagation. As required the total helicity, which is the volume integral of $h_{\tm{free}}$, acts as the generator of this transformation.

An important consequence of the Heaviside-Larmor symmetry is that all physical quantities must also be invariant under it \cite{Rose}. This means, for example, that the energy density $(\mathbf{E}\cdot \mathbf{D}+ \mathbf{B}\cdot \mathbf{H})/2$, and the kinetic and canonical momentum densities $\mathbf{E}\times\mathbf{H}$ and $\mathbf{D}\times\mathbf{B}$ \cite{jackson}, are satisfactory, but that $\mathbf{E}\times\mathbf{B}$ is not, and so lacks the physical significance of the other cross-products. This condition also applies to optical forces on a medium, and we note that the the Einstein-Laub force density \cite{einstein1908,BarnettLoudon2015}
\begin{equation}
\mathbf{f}=(\mathbf{P}\cdot\boldsymbol\nabla)\mathbf{E}+\frac{\partial\mathbf{P}}{\partial t} \times \mathbf{H}+ (\mathbf{M}\cdot\boldsymbol\nabla)\mathbf{H}-\frac{\partial\mathbf{M}}{\partial t}\times \mathbf{E},
\end{equation}
also satisfies the symmetry. Here, $\mathbf{P}$ and $\mathbf{M}$ are the polarisation and magnetisation densities within the medium.

\begin{figure}
\includegraphics{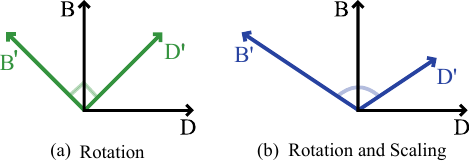}
\caption{Illustration comparing two ways of defining the duality transform in a medium. \textbf{(a)} The duality transform defined in equation \eqref{eq:dualityTransformDB} implies a simple rotation between the $\mathbf{D}$ and $\mathbf{B}$ fields; \textbf{(b)} The duality transforms defined in equations \eqref{eq:dualityTransformEpsilonMu} and \eqref{eq:dualityTransformRescaling} are equivalent to a rotation and scaling, and alter both the relative lengths of $\mathbf{D}'$ and $\mathbf{B'}$ and the angle between them. It is clear that the transformations \eqref{eq:dualityTransformEpsilonMu} and \eqref{eq:dualityTransformRescaling} are both incompatible with those in equation \eqref{eq:dualityTransformDB}.}
\label{fig:sketch}
\end{figure}

A problem arises when we seek the conserved quantity that is naturally associated with the Heaviside-Larmor transformation in a medium. If, for example, we consider the simplest constitutive relations, $\mathbf{D}=\varepsilon\mathbf{E}$ and $\mathbf{B}=\mu\mathbf{H}$, and impose the transformation from $\mathbf{E}$ and $\mathbf{H}$ to $\mathbf{E}'$ and $\mathbf{H}'$, then we find
\begin{align}\label{eq:dualityTransformEpsilonMu}
\mathbf{D}'&= \varepsilon\mathbf{E}'=\cos{\theta}\, \mathbf{D}+\sin{\theta}\,\left(\varepsilon/\mu\right)\mathbf{B},\nonumber\\
\mathbf{B}'&= \mu \mathbf{H}' = \cos{\theta}\,\mathbf{B}-\sin{\theta}\,\left(\mu/\varepsilon\right)\mathbf{D},
\end{align}
which is not of the form required by the Heaviside-Larmor symmetry. This problem is depicted for both $\mathbf{D'}$ and $\mathbf{B}'$ in Figure~\ref{fig:sketch}. The situation is worse in chiral media, for which $\mathbf{D}$ depends on both $\mathbf{E}$ and $\mathbf{H}$, while $\mathbf{B}$ is a function of both $\mathbf{H}$ and $\mathbf{E}$ \cite{Silverman1986,Lekner1996}. The transformations \eqref{eq:dualityTransformEpsilonMu} and \eqref{eq:dualityTransformDB} are compatible only when $\varepsilon/\mu = 1$. In this case, we say that the medium is dual-symmetric.

\section{Defining a Duality Transform}

One way in which the incompatibility of \eqref{eq:dualityTransformDB}, \eqref{eq:dualityTransformEpsilonMu} and the constitutive relations has been addressed previously is to define a duality transform in media which explicitly includes the permittivity and permeability \cite{Corbaton2013, JoergsDualityCond, noriHelicity, neel1}. That is, to use the transformation
\begin{align}\label{eq:dualityTransformRescaling}
\mathbf{E}' &= \cos{\theta}\, \mathbf{E}+\sin{\theta}\, \sqrt{\mu/\varepsilon}\,\mathbf{H},\nonumber\\
\mathbf{H}' &= \cos{\theta}\, \mathbf{H}-\sin{\theta}\,\sqrt{\varepsilon/\mu}\,\mathbf{E}, \nonumber\\
\mathbf{D}' &= \cos{\theta}\, \mathbf{D}+\sin{\theta}\, \sqrt{\varepsilon/\mu}\,\mathbf{B},\nonumber\\
\mathbf{B}' &= \cos{\theta}\, \mathbf{B}-\sin{\theta}\,\sqrt{\mu/\varepsilon}\,\mathbf{D},
\end{align}
in place of \eqref{eq:dualityTransformDB}. This transformation is a symmetry of Maxwell's equations \eqref{eq:MaxwellsFreeSpace} under the condition that $\grad{}\sqrt{\varepsilon/\mu}=\mb{0}$ throughout space. This approach leads to an optical helicity density similar to the free space expression \eqref{eq:helicityFreeSpace}, with the addition of appropriate factors of $\varepsilon$ and $\mu$ \cite{noriHelicity}. In chiral media, however, the situation is more complicated, and the helicity density derived from this transformation above includes extra contributions proportional to the energy density in the medium \cite{neel1}.

The difference between the transformations \eqref{eq:dualityTransformDB} and \eqref{eq:dualityTransformRescaling} can be emphasised by noting that \eqref{eq:dualityTransformDB} corresponds to a rotation between $\mathbf{D}$ and $\mathbf{B}$, while \eqref{eq:dualityTransformRescaling} corresponds to a rotation \textit{and scaling}, with a scale factor which depends on both $\varepsilon$ and $\mu$. Dispersion in the absence of absorption can be accommodated for in the transformation \eqref{eq:dualityTransformRescaling} \cite{noriHelicity}. From this point of view, however, treating dispersive media amounts to applying a different scaling for the different frequency components of the fields.

This approach works for linear, lossless materials. However, there are features of the transformation \eqref{eq:dualityTransformRescaling} that make it very difficult to extend to more general cases. It is unclear how to account for frequency-dependent absorption, which prohibits the treatment of realistic media that obey the Kramers-Kronig relations. Nonlinear responses are excluded from the outset, as the \textit{definition} of the transformation \eqref{eq:dualityTransformRescaling} assumes that the field pairs $\mathbf{D}/\mathbf{B}$ and $\mathbf{E}/\mathbf{H}$ are connected by linear constitutive relations.

Furthermore, we may note that the transformations \eqref{eq:dualityTransformRescaling} imply the following transformation of the polarisation $\mathbf{P}$, and magnetisation $\mathbf{M}$,
\begin{align}\label{eq:dualityTransformPMRescaling}
\mathbf{P}'=\cos{\theta} \mathbf{P}+\sin\theta \sqrt{\frac{\varepsilon}{\mu}} \mathbf{M}+\sin\theta \Bigg(\sqrt{\frac{\varepsilon}{\mu}}-\sqrt{\frac{\mu}{\varepsilon}}\Bigg)\mathbf{H},\nonumber\\
\mathbf{M}'=\cos{\theta} \mathbf{M} - \sin{\theta} \sqrt{\frac{\mu}{\varepsilon}} \mathbf{P}+\sin\theta \Bigg(\sqrt{\frac{\varepsilon}{\mu}}-\sqrt{\frac{\mu}{\varepsilon}}\Bigg)\mathbf{E}.
\end{align}
We see that, while the transformations \eqref{eq:dualityTransformRescaling} may be a symmetry of Maxwell's equations \eqref{eq:MaxwellsFreeSpace} even when $\varepsilon\neq\mu$, they then imply transformations of $\mathbf{P}$ and $\mathbf{M}$ that depend explicitly on the fields, rather than the simple rotation implied by \eqref{eq:dualityTransformDB}.\footnote{There are some cases in which the transformation (\ref{eq:dualityTransformPMRescaling}) can be viewed as providing a scaled rotation between $\mathbf{P}$ and $\mathbf{M}$. If we have $\mathbf{D}=\varepsilon\mathbf{E}$ and $\mathbf{B}=\mu\mathbf{H}$, then we can write $\mathbf{P}=\chi_e\mathbf{E}$ and $\mathbf{M}=\chi_m\mathbf{H}$ with $\varepsilon=1+\chi_e$ and $\mu=1+\chi_m$. Using these relationships in (\ref{eq:dualityTransformPMRescaling}), we obtain \begin{align}
\mathbf{P}'=\cos{\theta} \mathbf{P}+\sin\theta \sqrt{\frac{\varepsilon}{\mu}}\frac{\chi_e}{\chi_m} \mathbf{M},\nonumber\\
\mathbf{M}'=\cos{\theta} \mathbf{M} - \sin{\theta} \sqrt{\frac{\mu}{\varepsilon}} \frac{\chi_m}{\chi_e}\mathbf{P},\nonumber
\end{align} which, for a plane wave propagating through the medium, would effect a physical rotation of the $\mathbf{P}$ and $\mathbf{M}$ vectors induced by the wave about the direction of propagation. For more general constitutive relations -- for example, a chiral medium described by $\mathbf{D}=\epsilon(\mathbf{E} + \beta\boldsymbol{\nabla}\times\mathbf{E})$, $\mathbf{B}=\mu(\mathbf{H} + \beta\boldsymbol{\nabla}\times\mathbf{H}$) -- this interpretation is less clear, even when (\ref{eq:dualityTransformRescaling}) remains a symmetry.}

All of these issues stem from the fact that \eqref{eq:dualityTransformRescaling} explicitly includes the constitutive parameters $\varepsilon$ and $\mu$ in the transformation. Our approach here is to eschew \eqref{eq:dualityTransformRescaling}, and to instead consider
\begin{align}\label{eq:dualityTransformAC}
\mathbf{A}'=\cos\theta\,\mathbf{A}+\sin\theta\,\mathbf{C}, && \mathbf{C}'=\cos\theta\,\mathbf{C}-\sin\theta\,\mathbf{A},\nonumber\\
\mathbf{P}'=\cos\theta\,\mathbf{P}+\sin\theta\,\mathbf{M}, && \mathbf{M}'=\cos\theta\,\mathbf{M}-\sin\theta\,\mathbf{P},
\end{align}
as the duality transformation from which to derive the optical helicity inside a general medium. The key to this approach is that the fields $\mathbf{P}$ and $\mathbf{M}$ are not derived from the imposition of constitutive relations, but are explicitly included in an effective model of the material's internal degrees of freedom (as in \cite{hopfield,Huttner_1991}). 

The explicit inclusion of dynamical material degrees of freedom leads to the natural emergence of a matter component in the medium helicity. We can expect, therefore, the resulting helicity to have, generally, properties different from the case when only the photon degrees of freedom are retained (such as \eqref{eq:dualityTransformRescaling}), particularly for excitations that behave in a matter-like rather than photon-like manner. We emphasise that there are no \textit{a priori} reasons to expect the matter helicity to be closely tied to the circular polarisations of the $\mathbf{A}$, $\mathbf{P}$ or $\mathbf{M}$ fields. Nonetheless, we must also recover the familiar behaviour of the helicity being proportional to the photon-energy density for excitations that are photon-like. The approach taken here is the helicity counterpart of treating the electromagnetic and matter degrees of freedom on equal footing in the energy density, which is required for the optical Thomas-Reiche-Kuhn sum rules \cite{steveTRK} to be fulfilled.

\section{The Helicity Operator as a Generator}

In order to enforce both of the transformations in Eq.~\eqref{eq:dualityTransformAC} and, thereby, generate the Heaviside-Larmor transformation \eqref{eq:dualityTransformDB}, we need to extend the form of the helicity and of its flux to include \textit{both} field and matter components. This suggests that the total helicity density should take the form
\begin{equation}
h = \frac{1}{2}\left(\Ah\cdd\Bh-\Ch\cdd\Dh\right)+\Pih^P\cdd\Mh-\Pih^M\cdd\Ph, \label{eq:helicity}\\
\end{equation}
where $\Pih^P$ and $\Pih^M$ are the canonical momenta conjugate to $\mathbf{P}$ and $\mathbf{M}$ respectively. This helicity now includes an explicit contribution from the matter, as well as from the electromagnetic field. We base this form of the helicity density on the fact its volume integral generates the full Heaviside-Larmor transformation~\eqref{eq:dualityTransformDB}. The simplest method to confirm this is to promote our fields to quantum mechanical operators, where 
\begin{align}
    &[\widehat{A}_i(\bx),\widehat{\Pi}_j^A(\bx')] = [\widehat{A}_i(\bx),-\widehat{D}_j(\bx')] = i\delta_{ij}^\perp(\bx-\bx'), \\
    &[\widehat{P}_i(\bx),\widehat{\Pi}_j^P(\bx')] = i\delta_{ij}^\perp(\bx-\bx'), \\
    &[\widehat{M}_i(\bx),\widehat{\Pi}_j^M(\bx')] = i\delta_{ij}^\perp(\bx-\bx').
\end{align}
We then find
\begin{align}
&e^{i\thh \intr \hat{h}(\br)}\Ehh(\bx) e^{-i\thh \intr \hat{h}(\br)}= \Ehh(\bx) + \thh \Hhh(\bx),\label{eq:hGenTransformE}\\
&e^{i\thh \intr \hat{h}(\br)}\Hhh(\bx)e^{-i\thh \intr \hat{h}(\br)} = \Hhh(\bx) - \thh \Ehh(\bx)\label{eq:hGenTransformH},\\
&e^{i\thh \intr \hat{h}(\br)}\Phh(\bx) e^{-i\thh \intr \hat{h}(\br)}= \Phh(\bx) + \thh \Mhh(\bx),\label{eq:hGenTransformX}\\
&e^{i\thh \intr \hat{h}(\br)}\Mhh(\bx)e^{-i\thh \intr \hat{h}(\br)} = \Mhh(\bx) - \thh \Phh(\bx)\label{eq:hGenTransformY},
\end{align}
to first order in $\theta$ and where all fields are transverse. The infinitesimal transformations of $\mathbf{D}$ and $\mathbf{B}$ follow directly. 

We have taken all fields to be transverse, as we here wish to study the consequences of this definition of helicity in the simplest possible case, that of a homogeneous medium. However, we note that helicity density in Eq.~\eqref{eq:helicity} will always only generate the duality transform for the transverse component of the electromagnetic fields, as $\Dh$ and $\Bh$ are necessarily transverse. This is a consequence of the fact that Maxwell's equations \eqref{eq:MaxwellsFreeSpace} only specify the transverse components of the electromagnetic fields: $\mathbf{D}$ and $\mathbf{B}$ are transverse, and $\mathbf{E}$ and $\mathbf{H}$ appear only through their curls, and so the longitudinal parts do not enter into the dynamics. From the perspective of the helicity as a generator, it is technically possible to include the longitudinal components of the matter parts of the helicity density, i.e. the polarisation $\Ph$ and magnetisation $\Mh$ and associated momenta, and promote the respective commutators to full $\delta$-functions. This may be important for the extension to inhomogeneous media. This is, however, outside the scope of what we wish to study here and we leave it to future work. 

Finally, we also find that the flux density of our generalised helicity density is simply
\begin{equation}
\mb{v} = \frac{1}{2}\lel(\Eh\times\Ah+\Hh\times\Ch\rer), \label{eq:spin}
\end{equation}
which we recognise as the density of the electromagnetic spin. For situations in which the helicity is locally conserved, the helicity density and this flux satisfy the local conservation law
\begin{equation}
\frac{\partial h}{\partial t} + \boldsymbol{\nabla}\cdot\mathbf{v}=0,
\end{equation}
as may be derived directly from Noether's theorem. The details of this are given in Appendix \ref{appendix:NoethersTheorem}.

\section{Examples of helicity in media}

By explicitly including a matter component, we have obtained a generally applicable form for the helicity and established that it is indeed the conserved quantity associated with the full Heaviside-Larmor symmetry. It remains to examine how this quantity appears for a given medium and, moreover, to determine the physical forms of the two canonical momenta $\widehat{\mathbf{\Pi}}^P$ and $\widehat{\mathbf{\Pi}}^M$, which will depend on the mathematical model employed. To this end we introduce a model Lagrangian of a simple magnetodielectric medium.

For simplicity, we here consider here a space filled by a homogeneous lossless magnetodielectric, with a single electric and a single magnetic resonance. The polarisation and magnetisation fields satisfy the equations of motion
\begin{align}
\frac{\partial^2\mathbf{P}}{\partial t^2}+\omega_E^2 \mathbf{P}&=\alpha ^2\mathbf{E},\nonumber\\
\frac{\partial^2\mathbf{M}}{\partial t^2}+\omega_M^2 \mathbf{M}&=\beta^2 \mathbf{H},\label{eq:PandMEquationsofMotion}
\end{align}
where $\omega_E$ and $\omega_M$ are the resonance frequencies, and $\alpha$ and $\beta$ are constants. In frequency-space, it is clear that these equations imply the constitutive relations $\mathbf{D}=\varepsilon\mathbf{E}$ and $\mathbf{B}=\mu\mathbf{H}$, with
\begin{align} \label{eq:epsilonMu}
\varepsilon(\omega)=1+\frac{\alpha^2}{(\omega_E^2-\omega^2)}, && \mu(\omega)=1+\frac{\beta^2}{(\omega_M^2-\omega^2)}.
\end{align}

The equations \eqref{eq:PandMEquationsofMotion}, along with Maxwell's equations, may be obtained from the Lagrangian density
\begin{align}\label{eq:Lagrangian}
&\mathcal{L}=\frac{1}{2} [\mathbf{E}^2-\mathbf{B}^2]+\mathbf{P}\cdot\mathbf{E}+\mathbf{M}\cdot\mathbf{B}\\
&+\frac{1}{2\alpha^2} \big[(\partial_t\mathbf{P})^2-\omega_E^2 \mathbf{P}^2\big] + \frac{1}{2\beta^2} \big[(\partial_t\mathbf{M})^2-(\omega_M^2+\beta^2) \mathbf{M}^2\big],\nonumber
\end{align}
and we take this Lagrangian as a starting point from which to derive an expression for the optical helicity. To begin, we note that the Lagrangian is invariant under the duality transform \eqref{eq:dualityTransformAC} if (and only if) $\alpha=\beta$ and $\omega_E=\omega_M$. If we set $\alpha=\beta$ and $\omega_E=\omega_M=\omega_0$, then we can apply Noether's theorem to recover the helicity directly, as shown in Appendix \ref{appendix:NoethersTheorem}. Having established this definition in the dual-symmetric case, we may then apply it in the general case where $\omega_E\neq\omega_M$ or $\alpha\neq\beta$, and the quantity is no longer locally conserved.

For our chosen model Lagrangian, the canonical momenta corresponding to the polarisation and magnetisation are given by
\begin{align}
\mathbf{\Pi}^P\equiv\frac{\partial\mathcal{L}}{\partial\dot{\mathbf{P}}}=\partial_t\mathbf{P}/\alpha^2,\label{eq:PiP}\\
\mathbf{\Pi}^M\equiv\frac{\partial\mathcal{L}}{\partial\dot{\mathbf{M}}}=\partial_t\mathbf{M}/\beta^2.\label{eq:PiM}
\end{align}
These may therefore be interpreted as bound polarisation and magnetisation currents for the Lagrangian \eqref{eq:Lagrangian}. 

\section{Quantised helicity}

We turn next to a microscopic quantum description of the helicity. In free space, the total helicity can be expressed as the difference between the number of left- and right-handed circularly polarised photons \cite{ranada1996,Barnett2012,candlin},
\begin{align} \label{eq:freeSpaceHelicityPhotonDiffrence}
\int d^3r\, \hat{h}_\tm{free} = \int d^3k \sum_{\lambda = \pm} \lambda \;\adh_\la(\bk)\ah_\la(\bk),
\end{align}
where $\hat{a}_\lambda^\dagger(\mathbf{k})$ and $\hat{a}_\lambda(\mathbf{k})$ are the creation/annihilation operators for a photon of wavevector $\mathbf{k}$, and $\lambda=\pm 1$ labels the two circular polarisations. This is another way of stating that the helicity per photon equals $\pm$ the photon energy divided by its frequency, as the energy density per photon in some mode is simply $k\, \adh_\la(\bk)\ah_\la(\bk)$. We will find that the helicity operator defined by equation \eqref{eq:helicity} admits a similar interpretation in a dual-symmetric medium: it is equal to the difference in the number of left- and right-handed \textit{polaritons} in the system. In other words, a polariton carries helicity that equals $\pm$ the total energy density over the polariton frequency. We shall also see that the relationship is not quite so simple in a general medium, as the matter part of the helicity can contribute with a different sign to the electromagnetic part.

To begin, we diagonalise the system and examine the polariton modes. From the Lagrangian \eqref{eq:Lagrangian}, we obtain the Hamiltonian
\begin{align}\label{eq:NonDSHamiltonian}
\hat{H} = &\intk \sum_{\lambda = \pm} \; k \;\adh_\la(\bk)\ah_\la(\bk) \\
&+ \widetilde{\om}_E\bdh_\la(\bk)\bh_\la(\bk)+\widetilde{\om}_M\cdh_\la(\bk)\ch_\la(\bk) \nonumber \\
&-i\frac{\alpha}{2}\sqrt{\frac{k}{\widetilde{\om}_E}}\lel[\ah_\la(\bk)-\adh_\la(-\bk)\rer]\lel[\bh_\la(-\bk)+\bdh_\la(\bk)\rer]\nonumber\\
&-\la\frac{\beta}{2}\sqrt{\frac{k}{\widetilde{\om}_M}}\lel[\ah_\la(\bk)+\adh_\la(-\bk)\rer]\lel[\ch_\la(-\bk)+\cdh_\la(\bk)\rer].\nonumber
\end{align}
where $\hat{a}_\lambda(\mathbf{k})$, $\hat{b}_\lambda(\mathbf{k})$ and $\hat{c}_\lambda(\mathbf{k})$ are the annihilation operators corresponding to the fields $\mathbf{A}$, $\mathbf{P}$ and $\mathbf{M}$ respectively. We have also defined $\omt_E^2 = \om_E^2+\alpha^2$ and $\omt_M^2 = \om_M^2+\beta^2$. We define the circular polarisation vectors $\mb{e}_{\pm}$ with the convention that $\mb{e}_{\la}(\bk)\cdd\mb{e}_{\la'}^{*}(\bk)=\delta_{\la \la'}$, $\mb{e}_{\la}(\bk)\cdd\bk =0$ for all $\la$, and $\mb{e}_{\la}(-\bk)=-\mb{e}_{-\la}(\bk)$. Our continuum annihilation and creation operators satisfy the bosonic commutation relations $[\ah_\la(\bk),\adh_{\la'}(\bk')] = \delta_{\la\la'}\delta(\bk-\bk')$, with similar relations for $\bh_\la(\bk)$ and $\ch_\la(\bk)$, and all other commutators equal to zero. The explicit expansions of $\hat{\mathbf{A}}$, $\hat{\mathbf{P}}$ and $\hat{\mathbf{M}}$ used to obtain \eqref{eq:NonDSHamiltonian} from \eqref{eq:Lagrangian} are given in Appendix \ref{appendix:Hamiltonian}.

We consider first the helicity in a dual-symmetric medium, and show that the total helicity is naturally expressible in terms of the polariton modes of the system. As mentioned above, in a dual-symmetric medium, the helicity operator defined by equation \eqref{eq:helicity} is simply equal to the difference between the number of left- and right-handed polaritons in the system. We will then go on to show how this picture is modified in more realistic media.

\subsection{Helicity in dual symmetric media}\label{section:helicityInDSMedia}

\begin{figure}
\centering\includegraphics[width=0.45\textwidth]{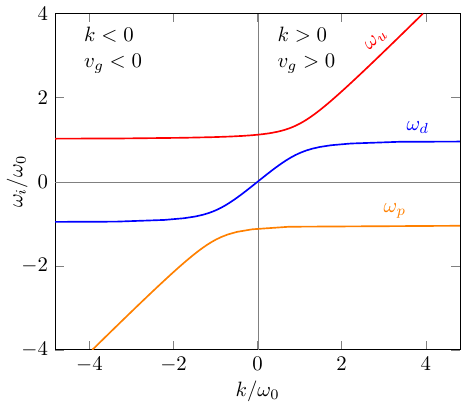}
\caption{The polariton branches, i.e. solutions to the dispersion relation in Eq.~\eqref{eq:dispersion}, with $\alpha = \omega_0/2$. The sign of $\omega_i$ has been chosen to ensure that the wave-vector and group velocity direction coincide.}
\label{fig:polaritons}
\end{figure}

In order to specialise (\ref{eq:NonDSHamiltonian}) to a dual-symmetric medium, we consider the case where $\omega_E=\omega_M\equiv\omega_0$ and $\alpha=\beta$. The Hamiltonian then becomes
\begin{align}\label{eq:Hamiltonian}
\hat{H} = \intk &\sum_{\lambda = \pm} \; k \;\adh_\la(\bk)\ah_\la(\bk) \\
&+ \widetilde{\om}_0\lel[\bdh_\la(\bk)\bh_\la(\bk)+\cdh_\la(\bk)\ch_\la(\bk)\rer]   \nonumber\\
&-i\kappa\lel[\ah_\la(\bk)-\adh_\la(-\bk)\rer]\lel[\bh_\la(-\bk)+\bdh_\la(\bk)\rer]\nonumber\\
&-\la\kappa\lel[\ah_\la(\bk)+\adh_\la(-\bk)\rer]\lel[\ch_\la(-\bk)+\cdh_\la(\bk)\rer].\nonumber
\end{align}
where we have defined the coupling parameter $\kappa = (\alpha/2)\sqrt{k/\omt_0}$. To diagonalise \eqref{eq:Hamiltonian}, we follow \cite{hopfield,kittel} and define polariton annihilation operators as linear combinations of the $\mathbf{A}$, $\mathbf{P}$ and $\mathbf{M}$ creation/annihilation operators
\begin{align}\label{eq:polariton}
    \ph_i(\bk,\la) = \al_i \ah_{\la}(\bk) &+ \altt_i \adh_{\la}(-\bk) + \be_i \bh_{\la}(\bk)+\bett_i \bdh_{\la}(-\bk) \nonumber\\
    &+ \ga_i \ch_{\la}(\bk)+\gatt_i \cdh_{\la}(-\bk),
\end{align}
with the requirement that $[\ph_i(\bk,\la),\hat{H}] = \om_i \ph_i(\bk,\la)$. The coefficients $\al_i$, $\altt_i$, $\be_i$, $\bett_i$, $\ga_i$ and $\gatt_i$ are constants, and the index $i$ labels the different polariton branches. The details of the diagonalisation are given in Appendix \ref{appendix:diagonalisation}. We find that the polariton branch frequencies, $\omega_i$, satisfy the dispersion relation
\begin{align}\label{eq:dispersion}
    k^2 = \om_i^2 n^2(\om_i) = \om_i^2\lel(1-\frac{\al^2}{\om_i^2-\om_0^2}\rer)^2.
\end{align}
The solutions to Eq.~\eqref{eq:dispersion} are plotted in Fig.~\ref{fig:polaritons}.

\begin{figure*}

\subfloat{
\centering\includegraphics[width=0.33\textwidth]{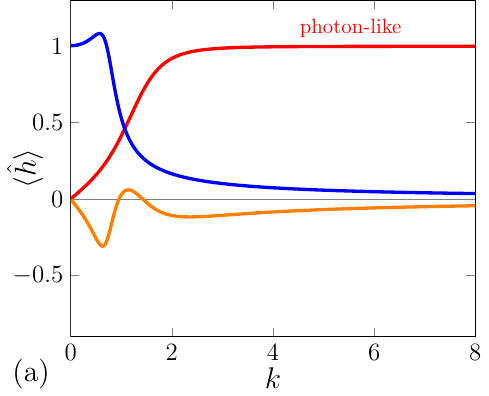}}
\subfloat{
\centering\includegraphics[width=0.33\textwidth]{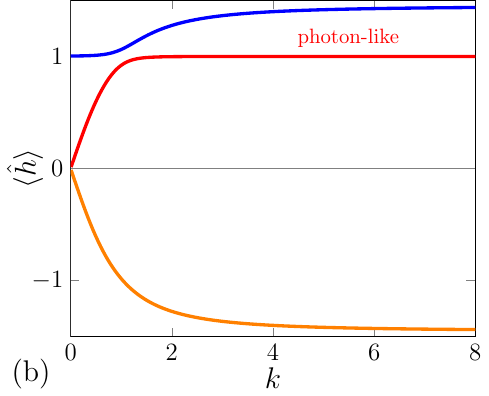}}
\subfloat{
\centering\includegraphics[width=0.33\textwidth]{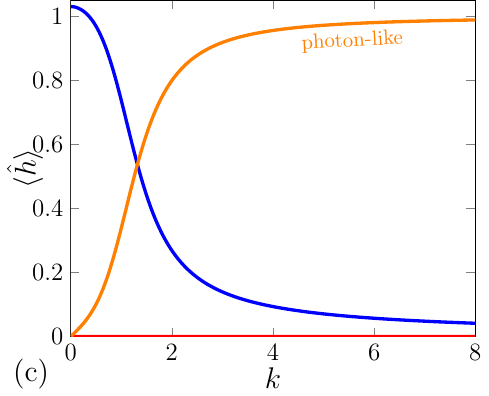}}
\\
\subfloat{
\centering\includegraphics[width=0.33\textwidth]{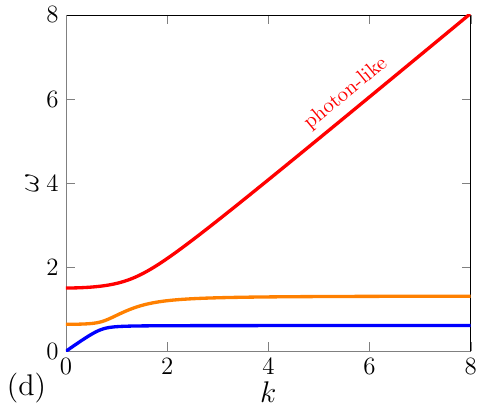}}
\subfloat{\centering\includegraphics[width=0.33\textwidth]{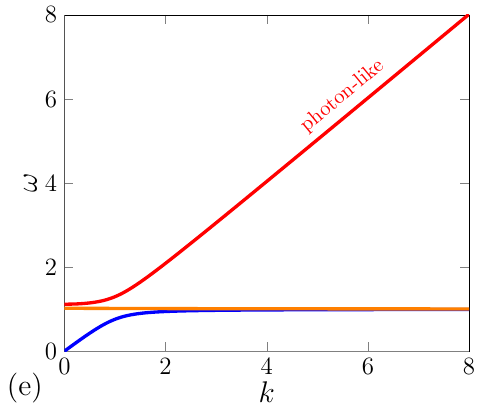}}
\subfloat{\centering\includegraphics[width=0.33\textwidth]{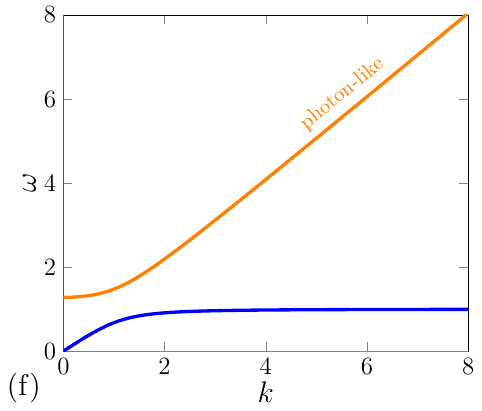}}

\caption{Some illustrative examples of the expectation value of $\hat{h}$ for an excitation in a single polariton mode, with a range of non-dual-symmetric model parameters. The corresponding dispersion curves for the polariton modes are shown below each graph. Figure 3a shows a typical case, with the helicity approaching unity when the branches are photon-like. Note that $\langle\hat{h}\rangle$ is not bounded between $\pm 1$, and the middle branch's contribution changes sign as $k$ varies. Figure 3b shows a case with $\omega_E=\omega_M$, but $\alpha\neq\beta$, where the asymptotic behaviour of the matter-like branches does not approach $0$ or $1$ for large $k$. Figure 3c shows a nonmagnetic material, which can be described in this formalism by letting $\omega_M \rightarrow \infty$. One of the branches then becomes a flat band at a very high frequency (not shown), and the system effectively reduces to two branches.}
\label{fig:helicityPerPolariton}
\end{figure*}

Once we have the explicit expressions for the polariton operators, we are in a position to express the helicity in terms of the polaritons. After some algebra, we obtain
\begin{align}\label{eq:helicityDiagonal}
:\intr{\hat{h}}: &= \intr :\frac{1}{2}\left(\Ahh\cdd\Bhh-\Chh\cdd\Dhh\right) \\ 
&\hspace{3cm} +\Pihh^P\cdd\Mhh-\Pihh^M\cdd\Phh : \nonumber\\ 
&=\sum_{\la = \pm} \intk \la \; \bigg[ \pdh_{u}(\bk,\la)\ph_{u}(\bk,\la) \\ \nonumber
&\hspace{3cm} + \pdh_{d}(\bk,\la)\ph_{d}(\bk,\la) \nonumber\\
&\hspace{3cm} - \pdh_{p}(\bk,\la)\ph_{p}(\bk,\la) \bigg],\nonumber
\end{align} 
where we have normal-ordered with respect to the polariton vacuum, and have labelled the three polariton branches $u$, $d$ and $p$. More details of this result's derivation are presented in Appendix \ref{appendix:diagonalHelicity}. Equation \eqref{eq:helicityDiagonal} is simply the difference between the number of positive and negative helicity polaritons, and is completely analogous to the free space expression \eqref{eq:freeSpaceHelicityPhotonDiffrence}. As polaritons are the eigenstates of the full Hamiltonian, the helicity equals $\pm$ the \textit{total} energy density divided by frequency, rather than just the \textit{electromagnetic} energy density. Note that one of the modes has a negative refractive index, and therefore contributes to the total helicity with opposite sign (as for this mode, the wave propagates in the opposite direction to the wave vector).

We note that the helicity is only expressible in this simple way when the medium is dual-symmetric (i.e. when $\alpha=\beta$ and $\omega_E=\omega_M$ such that $\varepsilon/\mu = 1$). This is to be expected, as the helicity is only expressible in terms of polariton number operators in a dual-symmetric medium. In a general lossless medium, the number of polaritons is conserved while the helicity is not. In such a medium, the total helicity operator can still be expressed in terms of circularly polarised polaritons, but includes more general products of polariton creation and annihilation operators, given in Appendix \ref{appendix:diagonalHelicity} (\ref{eq:helicityGeneralMediumPolaritons}). In the following subsection, we will discuss the helicity in such cases. The helicity and the Hamiltonian no longer commute, and we show that the expectation value of the helicity acquires an explicit time-dependence.

\subsection{Helicity in ordinary media}

The dual symmetric Hamiltonian considered in the previous section represents a very special case, where the helicity is exactly conserved. When $\alpha\neq\beta$ and/or $\omega_E\neq\omega_M$, the transformation (\ref{eq:dualityTransformAC}) is no longer a symmetry of the Hamiltonian (\ref{eq:NonDSHamiltonian}), and the helicity is no longer locally conserved. A consequence of this is that the commutator $[\hat{H}, \intr{\hat{h}}]\neq 0$, meaning that eigenstates of the Hamiltonian are not helicity eigenstates. Explicitly, the commutator is given by
\begin{widetext}
\begin{align}
\comm{\hat{H}}{\intr{\hat{h}}}&=
\sum_\lambda \frac{i}{2}\Bigg(\frac{\beta}{\alpha}\sqrt{\frac{\tilde{\omega}_E}{\tilde{\omega}_M}} + \frac{\alpha}{\beta}\sqrt{\frac{\tilde{\omega}_M}{\tilde{\omega}_E}}\Bigg)(\tilde{\omega}_E-\tilde{\omega}_M) \nonumber
\hat{b}_\lambda^\dagger(\mathbf{k})\hat{c}_\lambda(\mathbf{k})
 + \frac{i}{2}\Bigg(\frac{\beta}{\alpha}\sqrt{\frac{\tilde{\omega}_E}{\tilde{\omega}_M}}-\frac{\alpha}{\beta}\sqrt{\frac{\tilde{\omega}_M}{\tilde{\omega}_E}}\Bigg)(\tilde{\omega}_E+\tilde{\omega}_M)\hat{b}_\lambda(-\mathbf{k})\hat{c}_\lambda(\mathbf{k})\nonumber\\
&\hspace{0.5cm} + \frac{i\lambda}{2}\sqrt{\frac{k}{\tilde{\omega}_E}}(\alpha-\beta)\Big(-\hat{a}_\lambda(-\mathbf{k})\hat{b}_\lambda(\mathbf{k})-\hat{b}_\lambda(\mathbf{k})\hat{a}_\lambda^\dagger(\mathbf{k})\Big) + \frac{1}{2}\sqrt{\frac{k}{\tilde{\omega}_M}}(\alpha-\beta)\Big(\hat{a}_\lambda(-\mathbf{k})\hat{c}_\lambda(\mathbf{k})-\hat{a}_\lambda^\dagger(\mathbf{k})\hat{c}_\lambda(\mathbf{k})\Big) \nonumber\\
&\hspace{0.5cm} - \tm{H.c.},\label{eq:helicityHamiltonianCommutator}
\end{align}
\end{widetext}
which is generally non-zero. To arrive at this result, we have expressed $\hat{h}$ in terms of the $\hat{a}$, $\hat{b}$ and $\hat{c}$ operators, which can be found in Appendix \ref{appendix:diagonalHelicity} (\ref{eq:helicityGeneralMedium}). 

While the commutator \eqref{eq:helicityHamiltonianCommutator} is generally non-zero, it can be shown that its expectation value is equal to zero in the case of a polariton number state. For lossless media, polariton number states therefore exhibit a time-independent mean helicity. However, the value of this mean helicity is not generally $\pm 1$ for each circularly polarised polariton. This follows directly from the fact that we included the contribution associated with the matter in the helicity, which is not necessarily closely tied to the polariton circular polarisations. For a state with $n$ polaritons of wavevector $\mathbf{k}$ in branch $i$, we have
\begin{widetext}
\begin{align}
\intr &\bra{n(i,\mathbf{k},\lambda)} \hat{h}(\br)\ket{n(i,\mathbf{k},\lambda)}=\nonumber\\
&n\lambda(\alpha_i\alpha_i^*+\tilde\alpha_i\tilde\alpha_i^*)
+\Bigg(\frac{\beta}{\alpha}\sqrt{\frac{\tilde\omega_E}{\tilde\omega_M}}+\frac{\alpha}{\beta}\sqrt{\frac{\tilde\omega_M}{\tilde\omega_E}}\Bigg)\mathrm{Im}[\tilde\beta_i\tilde\gamma_i^*-\beta_i\gamma_i^*]
+\Bigg(\frac{\beta}{\alpha}\sqrt{\frac{\tilde\omega_E}{\tilde\omega_M}}-\frac{\alpha}{\beta}\sqrt{\frac{\tilde\omega_M}{\tilde\omega_E}}\Bigg)\mathrm{Im}[\beta_i\tilde\gamma_i^*-\tilde\beta_i\gamma_i^*],\label{eq:helicityPerPolariton}\end{align}
\end{widetext}
with no sum over the repeated index $i$ implied. Here, $\beta_i$, $\gamma_i$, $\tilde\beta_i$ and $\tilde\gamma_i$ now refer to the polariton expansion coefficients that diagonalise the general Hamiltonian \eqref{eq:NonDSHamiltonian}, rather than the dual-symmetric Hamiltonian \eqref{eq:Hamiltonian}. Explicit expressions for these are given in Appendix~\ref{appendix:generalDiagonalisation}. 

Plots of the expectation value (\ref{eq:helicityPerPolariton}) are shown in Figure~\ref{fig:helicityPerPolariton} for various combinations of model parameters, along with the corresponding polariton dispersion curves. Unlike in the dual-symmetric case, the total helicity of a polariton number state $\ket{n(i,\mathbf{k},\lambda)}$ is now $k$ dependent. It reaches a constant value only at large $k$, where we recover the free-space behaviour of one photon-like branch with unit helicity. This behaviour is to be expected, as the upper polariton branch will become increasingly photon-like at high $k$, where the helicity must be $\pm 1$ for the circular polarisations. In Figure~\ref{fig:helicityPerPolariton}a we show the helicity for an example medium where $\om_E \neq \om_M$ and $\alpha \neq \beta$, and we see that the branches typically have significant helicity only when they are photon-like.

When the resonant frequencies $\omega_E$ and $\omega_M$ are equal, illustrated in Figure~\ref{fig:helicityPerPolariton}b, we find that the photon-like branch still behaves in this manner. However, we also see that the matter-like branches carry significant helicity. Matching the resonant frequencies, but not the coupling strengths, therefore retains some features of the fully dual-symmetric case: the three branches asymptotically have constant, non-zero helicity. However, unlike the dual-symmetric case, the helicity reaches this value only asymptotically as $k$ increases. Furthermore, the asymptotic helicity values are not $\pm 1$ when $\alpha\neq\beta$. On a mathematical level, this is due to the factors of $\alpha/\beta$ and $\beta/\alpha$ in Eq.~\eqref{eq:helicityPerPolariton}. We therefore conclude that when the polarisation and magnetisation respond in phase, but at different strengths, enough symmetry is retained for the matter-like polaritons to carry helicity.

A special case of practical interest is the description of a non-magnetic medium. Although the formalism for defining helicity presented here requires that the material possesses both an electric and a magnetic response, a non-magnetic limit can be taken by letting $\omega_M \rightarrow \infty$, as shown in Figure \ref{fig:helicityPerPolariton}c. This effectively reduces the system to two branches, as one branch becomes a flat band at very high frequency. Nevertheless, the operator (\ref{eq:helicity}), obtained from the duality rotation (\ref{eq:dualityTransformAC}), may still be used to define the helicity. Note that for a polariton in the high frequency branch, $\langle\hat{h}\rangle \sim 0$.

\subsection{Helicity oscillations}
While the polariton number states considered above have a time-independent mean helicity, the helicity does generally depend on time. In this subsection, we explore the time-dependence of the helicity, and demonstrate that the average helicity is not always conserved through an explicit example. For this, we consider preparing a system in an eigenstate of helicity, and we calculate the expectation value of the total helicity as the state evolves in time. In order to construct a helicity eigenstate, we note that the helicity operator can be diagonalised by a similar procedure to that which we used to diagonalise the Hamiltonian in section \ref{section:helicityInDSMedia}. We write a linear combination of polariton creation annihilation operators,
\begin{align} \label{eq:helicityLoweringOperator}
\hat{\mathfrak{h}}(\bk,\lambda) = \sum_{i} a_i \ph_i(\bk,\lambda)+\tilde{a}_i \pdh_i(-\bk,\lambda)
\end{align}
and impose the condition that $[\int d^3r \, \hat{h},\hat{\mathfrak{h}}] = \Lambda \hat{\mathfrak{h}}$. The operator $\hat{\mathfrak{h}}$ may then be interpreted as an annihilation operator for a mode of well-defined helicity with helicity eigenvalue $\Lambda$. We note that the helicity eigenstates do not generally have eigenvalues of $\pm 1$, due to the addition of the matter component.

Helicity eigenstates can be explicitly constructed by applying the creation operator $\hat{\mathfrak{h}}^\dagger(\bk,\lambda)$ to the vacuum state. However, the helicity vacuum is generally squeezed with respect to the polariton vacuum (as the helicty vacuum is defined through a mixture of polariton annihilation and creation operators). This is difficult to realise experimentally, and we will therefore proceed by approximating the helicity eigenstate using only the polariton-number-conserving terms in the total helicity $\int d^3r \, \hat{h}$. This is equivalent to applying $\hat{\mathfrak{h}}^\dagger(\bk,\lambda)$ to the \textit{polariton} vacuum. In other words, we excite a mixture of all polariton branches at a single wavenumber. We note that this approximation does not impact the qualitative points that we wish to illustrate, and is a good approximation given that we are considering wavenumbers away from the medium resonances.

In this approximation, the helicity creation operator acting on the vacuum can be written $\hat{\mathfrak{h}}^\dagger(\bk,\lambda)\ket{0_\tm{polariton}} = \sum_i a_i \pdh(\bk,\lambda)\ket{0_\tm{polariton}}$, as $\ph_i\ket{0_\tm{polariton}} = 0$. This means that the terms proportional to $\hat{p}_i$ in $\hat{\mathfrak{h}}^\dagger(\bk,\lambda)$ (or equivalently, those proportional to $\hat{p}^\dagger_i$ in Eq.~\eqref{eq:helicityLoweringOperator}) will not contribute to the construction of a helicity eigenstate. We therefore see that we only need to construct an eigenstate of the number-conserving part of the helicity operator. The full helicity operator is given in terms of the polariton operators in Appendix \ref{appendix:diagonalHelicity} (\ref{eq:helicityGeneralMediumPolaritons}), and the number-conserving part of this is given by
\begin{align} \label{eq:numberConservingHelicity}
    \int d^3r \, \hat{h}^\tm{num}(t) = \intk \sum_{ij,\,\la} h_{ij}(\lambda)\pdh_i(\bk,\lambda)\ph_j(\bk,\lambda),
\end{align}
where we have introduced the shorthand
\begin{align}
&h_{ij}(\lambda) = \lambda \alpha_i \alpha^*_j + \lambda \tilde{\alpha}^*_j \tilde{\alpha}_i \\
&+ \frac{i}{2} \Big(\frac{\beta}{\alpha}\sqrt{\frac{\tilde\omega_E}{\tilde\omega_M}}+\frac{\alpha}{\beta}\sqrt{\frac{\tilde\omega_M}{\tilde\omega_E}} \Big)(\gamma^*_j\beta_i+\tilde\gamma_i\tilde\beta_j^*-\gamma_i\beta_j^*-\tilde\gamma_j^*\tilde\beta_i)\nonumber\\
&-\frac{i}{2}\Big(\frac{\beta}{\alpha}\sqrt{\frac{\tilde\omega_E}{\tilde\omega_M}}-\frac{\alpha}{\beta}\sqrt{\frac{\tilde\omega_M}{\tilde\omega_E}}\Big)(-\tilde\beta_i\gamma_j^*-\beta_j^*\tilde\gamma_i+\beta_i\tilde\gamma_j^*+\tilde\beta_j^*\gamma_i).\nonumber
\end{align}
The eigenvectors of the matrix $h_{ij}$ contain the required coefficients to write down a state $\ket{\psi}$, such that $\int d^3r \,\hat{h}\ket{\psi} = \Lambda \ket{\psi}$, in terms of a superposition of the polariton operators. This generally takes the form
\begin{align} \label{eq:helicityEigenstate}
\ket{\psi}=\lel[a_1\pdh_1(\bk,\lambda)+a_2\pdh_2(\bk,\lambda)+a_3\pdh_3(\bk,\lambda)\rer]\ket{0}, 
\end{align}
where $a_i$ are the coefficients for the different polariton branches, normalised such that $|a_1|^2+|a_2|^2+|a_3|^2 = 1$. As we shall see, the average total helicity will oscillate in time. For this, we should note that, by construction, the time-dependence of the polariton operators is $\ph_i(\bk,\lambda,t) = \ph_i(\bk,\lambda)\exp(-i\om_i t)$. The total helicity density therefore takes the form
\begin{align}
    \int d^3r \, \hat{h}(t) &= \intk \sum_{ij,\,\la} h_{ij}(\lambda)\pdh_i(\bk,\lambda)\ph_j(\bk,\lambda)e^{i\lel(\om_i-\om_j\rer)t}\nonumber\\
&+ \tm{(non-polariton-number-conserving terms)},
\end{align}
where we have not given the explicit form of the non-number conserving terms (\textit{i.e.} the terms which contain products of two creation or two annihilation operators) as they will vanish when the expectation value of $\hat{h}$ is taken.  We can explicitly write down the time-dependence of the helicity by taking the expectation value of the operator (\ref{eq:numberConservingHelicity}) in the state (\ref{eq:helicityEigenstate}). Defining phases $\phi_i$ for the coefficients in (\ref{eq:helicityEigenstate}), so that $a_i = |a_i|\exp(i\phi_i)$, we see that
\begin{widetext}
\begin{align}
    \int d^3r \,\bra{\psi}\hat{h}(t)\ket{\psi} &= \sum_{\lambda'}\intkp \sum_{ij} h_{ij}(\lambda')e^{i\lel(\om_i-\om_j\rer)t}\delta^3(\bk'-\bk)\delta_{\lambda\lambda'}\lel[a_1^*\delta_{i1} +a_2^*\delta_{i2} +a_3^*\delta_{i3}\rer]\lel[a_1\delta_{j1} +a_2\delta_{j2} +a_3\delta_{j3}\rer] \nonumber\\
    &= \bigg\{|a_1|^2h_{11} + |a_2|^2h_{22} + |a_3|^2h_{33} + 2|h_{12}||a_1||a_2|\cos\lel[\lel(\om_2-\om_1\rer)t + \Delta\phi_{12}\rer] \\
    &\hspace{2cm} +  2|h_{13}||a_1||a_3|\cos\lel[\lel(\om_3-\om_1\rer)t+\Delta\phi_{13}\rer] + 2|h_{23}||a_2||a_3|\cos\lel[\lel(\om_3-\om_2\rer)t+\Delta\phi_{23}\rer]\bigg\},\nonumber
\end{align}
\end{widetext}
where we have used that $h_{ij} = h_{ji}^*$ in the final step, which follows from the fact that $\hat{h}(t)$ is Hermitian, and we have defined $\Delta\phi_{ij} \equiv \phi_i-\phi_j$. It is evident that the average helicity oscillates in time at the difference frequency between the branches. Only in a dual-symmetric medium are $h_{12}$, $h_{13}$, and $h_{23}$ identically zero, leading to a time-independent average helicity. 

An analogy can be drawn between the oscillatory behaviour of the helicity eigenstates shown above, and the phenomenon of neutrino oscillations \cite{neutrinoOscillations}. A beam of neutrinos prepared in an initially pure flavour eigenstate will be found, upon propagation, to contain a mixture of flavours, as the flavour eigenstates are not eigenstates of the full Hamiltonian \cite{Bigi2009}. In our case, a similar effect occurs for a system initially prepared in a pure helicity eigenstate, likewise stemming from the fact that the total helicity and the Hamiltonian do not commute. This phenomenon occurs whenever there is a superposition of different polariton branches at the same wavevector $\bk$. We note that, in practice, creating this kind of superposition state requires the excitation of \textit{both} matter-like and photon-like polariton modes.

\section{Discussion and conclusion}
In this article, we set out to find an expression for the optical helicity which is valid within any optical medium, and to resolve the mutual incompatibility between the duality transform \eqref{eq:dualityTransformDB}, and the constitutive equations. We approached this by using the duality transform \eqref{eq:dualityTransformAC} as the starting point from which to define the optical helicity, which provides the appropriate rotation between $\mathbf{E}$ and $\mathbf{H}$, and also $\mathbf{D}$ and $\mathbf{B}$, without coming into conflict with constitutive relations or presupposing any particular properties of the medium. This works because, in our approach, the constitutive equations connecting the fields emerge as a consequence of the Lagrangian and the associated equations of motion, rather than being enforced from the outset. We therefore require a description of the material's internal degrees of freedom in the Lagrangian, and this explicit description of the material causes a medium contribution to the helicity to naturally emerge. 

We find that the helicity defined in this way retains many properties that we are familiar with, but differences arise as this helicity also contains the twist associated with the matter. As polaritons are the natural quasi-particle excitations in optical media, we have used them to frame our discussions. In all cases, when the polariton is photon-like, it retains the interpretation of carrying $\pm$ the energy density over frequency. However, it is generally the \textit{total} energy that is important, rather than only the \textit{electromagnetic} energy. We find that, in a dual-symmetric medium where $\varepsilon = \mu$, the helicity per polariton is always $\pm$ the \textit{total} energy density over polariton frequency. However, in more general optical media, the helicity is not conserved, leading to a $k$-varying helicity for single polariton-branch excitations. Broadly speaking, the $k$-dependence can be seen as a consequence of the fact that, unlike in free space or in a dual-symmetric medium, the field vectors for circularly polarised modes are not generally eigenstates of the duality transformation; this point of view is outlined mathematically in Appendix \ref{appendix:dualityTransformEigenvalues}. The precise details of the $k$-dependence depend on the polariton character, and the details of the polarisation and magnetisation response. Interestingly, this non-conservation also leads to the possibility of helicity oscillations, akin to neutrino oscillations.
 
When comparing our results to those obtained elsewhere in the literature, it is important to appreciate that \eqref{eq:dualityTransformRescaling} and \eqref{eq:dualityTransformAC} are, in general, distinct symmetry transformations, and so one would not expect the conserved quantities associated with these to coincide exactly. Even in the case where $\varepsilon=\mu$, in which \eqref{eq:dualityTransformRescaling} reduces to \eqref{eq:dualityTransformAC}, we note that the helicity arising from Eq.~\eqref{eq:dualityTransformAC} contains a matter contribution absent from the helicity derived from Eq.~\eqref{eq:dualityTransformRescaling}. This situation is analogous to that discussed in Ref.~\cite{Huttner_1991} which points out that the total electromagnetic energy used in phenomenological quantisation schemes differs from the total energy of the full system by a matter component. Both are, of course, valid notions of helicity. We argue, however, that the approach taken here is superior as it explicitly takes the contribution from the matter into account. A somewhat different way of defining a ``matter'' contribution to the helicity has been presented by Fernandez-Corbaton \cite{staticMagneticHelicity}, where the static, zero-frequency contribution to the helicity in the presence of magnetic dipole sources is interpreted as helicity stored by the matter. 

The electromagnetic helicity and optical chirality \cite{Tang2010} are closely related quantities, and it has been proposed to generalise the optical chirality to dispersive, lossy media by taking $\frac{1}{2}\Big(\mathbf{E}\times(\boldsymbol{\nabla}\times\mathbf{H})+\mathbf{H}\times(\boldsymbol{\nabla}\times\mathbf{E})\Big)$ as the chirality flux in a medium \cite{VasquezLozano2019}. For monochromatic fields in free space, the chirality density is equal to $\omega^2$ times the helicity density, and likewise for the fluxes, and it might be supposed that a similar relationship should hold for monochromatic fields in a lossless medium. A similar relationship between the fluxes does indeed hold, but only for a different possible definition of the chirality flux: our helicity flux, $\mathbf{v}$, is given by (\ref{eq:spin}), and so a chirality flux that is equal to $(\omega^2/v_p^2)\mathbf{v}$ for monochromatic fields would be $\frac{1}{2}\Big(\mathbf{E}\times(\boldsymbol{\nabla}\times\mathbf{B})+\mathbf{H}\times(\boldsymbol{\nabla}\times\mathbf{D})\Big)$.

The extension of the Lagrangian \eqref{eq:Lagrangian} to allow for multiple resonances in $\varepsilon$ and $\mu$ may be achieved by straightforward modification of the harmonic oscillator terms. Lossy media can be treated by coupling the $\mathbf{P}$ and $\mathbf{M}$ fields to a continuous reservoir of harmonic oscillators \cite{Huttner_1992, steves,BahtSipe2006, scheel1,Kheirandish2008,
philbinQuantisation}. Finally, chiral and magnetoelectric responses can also be added to these models by introducing additional terms into $\mathcal{L}$ which couple the $\mathbf{P}$ and $\mathbf{M}$ degrees of freedom \cite{Horsely2011}. Such changes will generally alter the expressions for the conjugate momenta $\mathbf{\Pi}^P$ and $\mathbf{\Pi}^M$ (\ref{eq:PiP}-\ref{eq:PiM}), and their interpretation. However, any such modifications will not affect the basic results \eqref{eq:helicity} and \eqref{eq:spin}.

A point worth mentioning is that the definition presented here relies on the ability to use a model for the medium that can be made symmetrical under the duality transform \eqref{eq:dualityTransformAC}, for at least some choice of model parameters. This requires that the polarisation and magnetisation can be written as functions of $\mathbf{D}$ and $\mathbf{B}$, or $\mathbf{E}$ and $\mathbf{H}$, but not some combination of the two. For example, a medium where $\mathbf{P}=\mathbf{P}(\mathbf{E},\mathbf{H})$ and $\mathbf{M}=\mathbf{M}(\mathbf{H},\mathbf{E})$ would fulfil this requirement, but one where $\mathbf{P}=\mathbf{P}(\mathbf{E},\mathbf{B})$ and $\mathbf{M}=\mathbf{M}(\mathbf{B},\mathbf{E})$ would not. It is possible to reformulate the description of such materials in a way that can be made symmetric under the duality transform \eqref{eq:dualityTransformAC}; we leave this to future work.

\begin{acknowledgments}
This research was supported by the Royal Commission for the Exhibition of 1851, the Royal Society (grant numbers RSRP/RE/220013, RP/150122 and RSRP/R/210005), the EPSRC (grant number EP/V048449/1) and the Leverhulme Trust. 
\end{acknowledgments}

%

\clearpage

\onecolumngrid

\begin{center}
    \Large Appendices
\end{center}
\appendix
\section{Noether's Theorem}\label{appendix:NoethersTheorem}
\noindent Here we provide the details of the application of Noether's theorem \cite{Noether1918} to derive an 
expression for the optical helicity from the Lagrangian \eqref{eq:Lagrangian}. Under the infinitesimal version of duality transformation \eqref{eq:dualityTransformAC},
\begin{align}
\mathbf{A}'=\mathbf{A}+\delta\theta\mathbf{C}, && \mathbf{C}'=\mathbf{C}-\delta\theta\mathbf{A},\nonumber\\
\mathbf{P}'=\mathbf{P}+\delta\theta\mathbf{M}, && \mathbf{M}'=\mathbf{M}-\delta\theta\mathbf{P},\label{eq:infinitesimalDualityTransform}
\end{align}
the Lagrangian becomes $\mathcal{L}'=\mathcal{L}(\mathbf{A'},\mathbf{C'},\mathbf{P'},\mathbf{M'})$. We first consider the change in the Lagrangian, $\delta \mathcal{L}\equiv\mathcal{L}'-\mathcal{L}$. Substituting the duality transformation into the Lagrangian \eqref{eq:Lagrangian}, and setting $\alpha=\beta$ and $\omega_E=\omega_M$, we find that
\begin{align}\label{eq:boundaryTerm}
\delta \cL &= \delta\thh\lel(\Eh\cdd\Bh+\Hh\cdd\Dh\rer) \\
&= \frac{\delta\thh}{2}\left(-\partial_t\left[\Ah\cdd\Bh+\Ch\cdd\Dh\right]-\dive{}\left[\Eh\times\Ah-\Hh\times\Ch\right]\right),\nonumber
\end{align}
which is a four-divergence, showing that the equations of motion remain unchanged under the transform \cite{NoetherBook, noetherReview}.

We then calculate $\delta S = \int_{t_A}^{t_B} dt \;\delta \mathcal{L}$ through variation of the fields. The result is given by
\begin{align}
\delta S &= \delta\thh \intr \int_{t_A}^{t_B} dt \left(\frac{\partial}{\partial t}\left[-\Ch\cdd\Dh+\Pih^P\cdd\Mh-\Pih^M\cdd\Ph\right] - \dive{}\left[\Hh\times\Ch\right]\right), \label{eq:noetherDerivHelicity}
\end{align}
where we have used that $\Pih^P = \partial_t \Ph/\alpha^2$ and $\Pih^M = \partial_t \Mh/\alpha^2$ in the chosen gauge and coupling. Equating the two expressions for $\delta \mathcal{L}$ from  Eq.~\eqref{eq:boundaryTerm} and Eq.~\eqref{eq:noetherDerivHelicity} leads to the local conservation law 
\begin{align}
\frac{\partial}{\partial t}\bigg[\frac{1}{2}\bigg(\Ah\cdd\Bh &- \Ch\cdd\Dh\bigg)+\Pih^P\cdd\Mh-\Pih^M\cdd\Ph\bigg] \nonumber\\
&+ \dive{}\left[\frac{1}{2}\lel(\Eh\times\Ah+\Hh\times\Ch\rer)\right] = 0,
\end{align}
which explicitly shows the helicity density \eqref{eq:helicity} and flux \eqref{eq:spin} from the main text.

\section{Hamiltonian in field variables}\label{appendix:Hamiltonian}
In terms of $\mathbf{A}$, $\mathbf{P}$ and $\mathbf{M}$, and their conjugate variables, the Hamiltonian obtained from the Lagrangian in Eq.~\eqref{eq:Lagrangian} takes the form
\begin{align}
\hat{H} &= \intr \frac{1}{2}\lel[\left(\Pihh^A\right)^2+\Bhh^2\rer]\nonumber + \frac{1}{2\alpha^2}\lel[\alpha^4\left(\Pihh^P\right)^2+\omt_E^2\Phh^2\rer] + \frac{1}{2\beta^2}\lel[\beta^4\left(\Pihh^M\right)^2+\omt_M^2\Mhh^2\rer] \nonumber\\
&\hspace{1.5cm} +\Phh\cdd\Pihh^A-\Mhh\cdd\Bhh,\nonumber
\end{align}
If we expand the fields as
\begin{align}\label{eq:AXYexpansion}
    \Ahh &=\sum_{\la=\pm}\intkf \frac{1}{\sqrt{2k}}\bigg[\mb{e}_\la(\bk) \ah_\la(\bk)e^{i\bk\cdd\bx} + \mb{e}_{-\la}(\bk) \adh_\la(\bk)e^{-i\bk\cdd\bx} \bigg], \nonumber\\
    \Phh &=\sum_{\la=\pm}\intkf \frac{\alpha}{\sqrt{2\omt_E}}\bigg[\mb{e}_\la(\bk) \bh_\la(\bk)e^{i\bk\cdd\bx} + \mb{e}_{-\la}(\bk) \bdh_\la(\bk)e^{-i\bk\cdd\bx} \bigg], \\
    \Mhh &=\sum_{\la=\pm}\intkf \frac{\beta}{\sqrt{2\omt_M}}\bigg[\mb{e}_\la(\bk) \ch_\la(\bk)e^{i\bk\cdd\bx}  + \mb{e}_{-\la}(\bk) \cdh_\la(\bk)e^{-i\bk\cdd\bx} \bigg], \nonumber
\end{align}
then we arrive at the Hamiltonian seen in the main text~\eqref{eq:NonDSHamiltonian}. Making the substitutions $\beta=\alpha$ and $\omega_E=\omega_M\equiv\omega_0$ leads to~\eqref{eq:Hamiltonian}. We note that this latter Hamiltonian is dual symmetric with respect to the transform in $(\Dhh,\Bhh)$, as $\Pihh^A = -\Dhh$.

\section{Details of the diagonalisation}\label{appendix:diagonalisation}
Here we present the details of the diagonalisation of \eqref{eq:Hamiltonian}, leading to expressions for the Hopfield coefficients appearing in \eqref{eq:polariton}. We begin with some preparatory results, which will be useful in expanding the helicity operator to obtain Eq.~\eqref{eq:helicityDiagonal} . First, we can invert Eq.~\eqref{eq:polariton} by noting that, as the polariton operators form a complete basis set, we must have that $\ah_\la(\bk) = \sum_i f_i \ph_i(\bk,\la)+g_i\pdh_i(-\bk,\la)$ for some constants $f_i$ and $g_i$ (and similarly for the other operators). Computing the commutators $\lel[\ah_\la(\bk'),\ph_i(\bk,\la)\rer]$ and $\lel[\ah_\la(\bk'),\pdh_i(-\bk,\la)\rer]$ allows us to identify the constants,
\begin{align}
\ah_\la(\bk) &= \sum_i \al_i^*\ph_i(\bk,\la)-\altt_i\pdh_i(-\bk,\la), \nonumber   \\
\bh_\la(\bk) &= \sum_i \be_i^*\ph_i(\bk,\la)-\bett_i\pdh_i(-\bk,\la), \label{eq:Invert}   \\
\ch_\la(\bk) &= \sum_i \ga_i^*\ph_i(\bk,\la)-\gatt_i\pdh_i(-\bk,\la), \nonumber
\end{align}

Furthermore, inserting the expressions in Eqns.~\eqref{eq:Invert} into the commutators $\lel[\ah_\la(\bk),\adh_{\la'}(\bk')\rer]=\delta_{\la \la'}\delta(\bk-\bk')$, etc., produces the relations
\begin{align}
\sum_i |\al_i|^2-|\altt_i|^2 &= 1,\nonumber\\
\sum_i |\be_i|^2-|\bett_i|^2 &= 1,\label{eq:sum_rules}\\
\sum_i |\ga_i|^2-|\gatt_i|^2 &= 1,\nonumber
\end{align}
which are types of sum rules. These sum rules are important, and we must always make sure that they are respected in order to have proper commutators. 

We now derive the explicit expressions for the Hopfield coefficients. Using $\lel[\ph_i(\bk,\la),\hat{H}\rer] = \om_i \ph_i(\bk,\la)$ \cite{hopfield}, along with the commutators for the creation and annihilation operators, implies the algebraic relations
\begin{align}\label{eq:algebraicRelations}
    (\om_i-k)\al_i &= \kappa\lel[-i(\be_i-\bett_i)-\la(\ga_i-\gatt_i)\rer], \nonumber\\
    (\om_i+k)\altt_i &= \kappa\lel[i(\be_i-\bett_i)-\la(\ga_i-\gatt_i)\rer], \nonumber\\    
    \be_i &= i\kappa\lel(\frac{\al_i+\altt_i}{\om_i-\omt_0}\rer), \nonumber\\
    \bett_i &= i\kappa\lel(\frac{\al_i+\altt_i}{\om_i+\omt_0}\rer), \\
    \ga_i &= -\la\kappa\lel(\frac{\al_i-\altt_i}{\om_i-\omt_0}\rer), \nonumber\\
    \gatt_i &= -\la\kappa\lel(\frac{\al_i-\altt_i}{\om_i+\omt_0}\rer). \nonumber
\end{align}
Along with this, we also get a normalisation from the the commutator $\lel[\ph_i(\bk,\la),\pdh_j(\bk',\la')\rer] = \delta_{ij}\delta_{\la \la'}\delta(\bk-\bk')$:
\begin{align}\label{eq:HopfieldNormalisation}
|\al_i|^2-|\altt_i|^2+|\be_i|^2-|\bett_i|^2+|\ga_i|^2-|\gatt_i|^2 = 1.
\end{align}
This completely solves the system, and leads to explicit expressions for our coefficients $\al_i$, $\altt_i$, $\be_i$, $\bett_i$, $\ga_i$ and $\gatt_i$.
It is important to note, however, that one of the normal mode frequencies defined through Eq.~\eqref{eq:dispersion} has a negative group velocity. In fact, if we simplify Eq.~\eqref{eq:dispersion} and pick the positive part of the square root such that 
\begin{align} \label{eq:dispersionRelationOne}
n(\om_i) = \frac{k}{\om_i} = \lel(1-\frac{\al^2}{\om_i^2-\om_0^2}\rer),
\end{align}
then we find two positive frequency modes and one negative. If we on the other hand choose the negative square root, we arrive at
\begin{align} \label{eq:dispersionRelationTwo}
n(\om_i) = \frac{k}{\om_i} = -\lel(1-\frac{\al^2}{\om_i^2-\om_0^2}\rer),
\end{align}
along with two negative frequency and one positive frequency mode. We want to choose the positive frequency (positive norm) modes, and notation simplifies if we define 
\begin{align}
n_{u/d}(\om_{u/d}) = 1-\frac{\al^2}{\om_{u/d}^2-\om_0^2};\hspace{0.1cm} n_p(\om_p) = -1+\frac{\al^2}{\om_p^2-\om_0^2}\nonumber,
\end{align}
where we have chosen the modes $\om_i = \{\om_u,\om_p,\om_d\}$ in descending order of oscillation frequency.
Manipulating the algebratic relations in Eqns.~\eqref{eq:algebraicRelations} yields
\begin{align}
\lel(\frac{\om_i}{k}-1\rer)\al_i+\lel(\frac{\om_i}{k}+1\rer)\altt_i = (\al_i-\altt_i)\frac{ \al^2}{\om_i^2-\omt_0^2}, \nonumber   
\end{align}
at which stage we must distinguish between $\om_{u/d}$ and $\om_p$. For $\om_{u/d}$ we have
\begin{align}
\lel(\frac{\om_{u/d}}{k}-1\rer)\al_{u/d} &+ \lel(\frac{\om_{u/d}}{k}+1\rer)\altt_{u/d} \nonumber\\
& = \lel(\al_{u/d}-\altt_{u/d}\rer)\lel(\frac{\om_{u/d}}{k}-1\rer) \nonumber\\
& \Rightarrow \altt_{u/d} = 0,    \nonumber
\end{align}
whereas for $\om_p$ we find that
\begin{align}
\lel(\frac{\om_p}{k}-1\rer)\al_p &+ \lel(\frac{\om_p}{k}+1\rer)\altt_p \nonumber\\
&= (-\al_p+\altt_p)\lel(\frac{\om_p}{k}+1\rer) \nonumber\\
&\Rightarrow \al_p = 0.    \nonumber
\end{align}
It is now straightforward to show that
\begin{align}
\lel|\al_{u/d}\rer|^2 &= \lel(1+2\kappa^2\lel[\frac{1}{(\om_{u/d}-\omt_0)^2}-\frac{1}{(\om_{u/d}+\omt_0)^2}\rer]\rer)^{-1} \label{eq:alpm}\\
\lel|\altt_p\rer|^2 &= \lel(-1+2\kappa^2\lel[\frac{1}{(\om_p-\omt_0)^2}-\frac{1}{(\om_p+\omt_0)^2}\rer]\rer)^{-1} \label{eq:alttp}
\end{align}
using Eqns.~\eqref{eq:algebraicRelations} and the normalisation condition Eq.~\eqref{eq:HopfieldNormalisation}. Importantly, we can note that this satisfies
\begin{align}
\lel|\al_u\rer|^2 + \lel|\al_d\rer|^2-\lel|\altt_p\rer|^2 = 1, \nonumber
\end{align}
which must be the case in order to preserve the canonical commutation relations. The phases of $\al_{u/d}$ and $\altt_p$ can be chosen freely, and the remaining Hopfield coefficients follow from Eqns.~\eqref{eq:algebraicRelations}.

\section{Diagonal form of the helicity}\label{appendix:diagonalHelicity}
In this section, we provide details of how to obtain Eq.~\eqref{eq:helicityDiagonal} in the main text, using the polariton coefficients obtained in the previous section. Let us express the total helicity, Eq.~\eqref{eq:helicity}, in terms of the photon and matter annihilation and creation operators, yielding:
\begin{align} \label{eq:helicityGeneralMedium}
\int{d^3r:\hat{h}:} &= \intr \lel[\frac{1}{2}\left(\Ahh\cdd\Bhh-\Chh\cdd\Dhh\right)+\Pihh^P\cdd\Mhh-\Pihh^M\cdd\Phh\rer] \nonumber\\
&= \sum_{\la = \pm}\intk \la\; \adh_{\la}(\bk)\ah_{\la}(\bk) \nonumber\\
&\hspace{1cm} + \frac{i}{2}\Bigg(\frac{\beta}{\alpha}\sqrt{\frac{\tilde{\omega_E}}{\tilde{\omega}_M}}+\frac{\alpha}{\beta}\sqrt{\frac{\tilde{\omega_M}}{\tilde{\omega_E}}}\Bigg)\left(\ch_{\la}(\bk)\bdh_{\la}(\bk)-\cdh_{\la}(\bk)\bh_{\la}(\bk)\right)\nonumber\\
&\hspace{1cm} - \frac{i}{2}\Bigg(\frac{\beta}{\alpha}\sqrt{\frac{\tilde{\omega_E}}{\tilde{\omega}_M}}-\frac{\alpha}{\beta}\sqrt{\frac{\tilde{\omega_M}}{\tilde{\omega_E}}}\Bigg)\left(\bh_{\la}(-\bk)\ch_{\la}(\bk)-\bdh_{\la}(-\bk)\cdh_{\la}(\bk)\right).
\end{align}
This can further be rewritten in terms of polaritons using Eqns.~\eqref{eq:Invert}, where we should note that we need to normal order the operators with respect to the polariton vacuum. This yields
\begin{align} \label{eq:helicityGeneralMediumPolaritons}
\int{d^3 r:\hat{h}:}=\int \, d^3 k \sum_\lambda \sum_{ij} &\hat{p}^\dagger_i (\mathbf{k},\lambda) \hat{p}_j(\mathbf{k},\lambda)\Bigg(\lambda \alpha_i \alpha^*_j + \lambda \tilde{\alpha}^*_j \tilde{\alpha}_i + \frac{i}{2} \Big(\frac{\beta}{\alpha}\sqrt{\frac{\tilde\omega_E}{\tilde\omega_M}}+\frac{\alpha}{\beta}\sqrt{\frac{\tilde\omega_M}{\tilde\omega_E}} \Big)(\gamma^*_j\beta_i+\tilde\gamma_i\tilde\beta_j^*-\gamma_i\beta_j^*-\tilde\gamma_j^*\tilde\beta_i) \nonumber\\
&-\frac{i}{2}\Big(\frac{\beta}{\alpha}\sqrt{\frac{\tilde\omega_E}{\tilde\omega_M}}-\frac{\alpha}{\beta}\sqrt{\frac{\tilde\omega_M}{\tilde\omega_E}}\Big)(-\tilde\beta_i\gamma_j^*-\beta_j^*\tilde\gamma_i+\beta_i\tilde\gamma_j^*+\tilde\beta_j^*\gamma_i)\Bigg)\nonumber\\
+\Bigg[&\hat{p}_i(-\mathbf{k},\lambda)\hat{p}_j(\mathbf{k},\lambda)\Big(-\frac{\lambda}{2}(\tilde\alpha_i^*\alpha_j^*+\alpha_i^*\tilde\alpha_j^*) + \frac{i}{2}\Big(\frac{\beta}{\alpha}\sqrt{\frac{\tilde\omega_E}{\tilde\omega_M}}+\frac{\alpha}{\beta}\sqrt{\frac{\tilde\omega_M}{\tilde\omega_E}}\Big)(-\gamma_i^*\tilde\beta_j^*+\tilde\gamma_i^*\beta_j^*)\nonumber\\
-&\frac{i}{2}\Big(\frac{\beta}{\alpha}\sqrt{\frac{\tilde\omega_E}{\tilde\omega_M}}-\frac{\alpha}{\beta}\sqrt{\frac{\tilde\omega_M}{\tilde\omega_E}}\Big)(\gamma_i^*\beta_j^*-\tilde\gamma_i^*\tilde\beta_j^*) \Big)+\mathrm{h.c}\Bigg].
\end{align}
In the dual-symmetric case, this simplifies to
\begin{equation}\label{eq:normalOrderedHelicity}
\begin{split}
\int{d^3 r:\hat{h}_{D. S.}:}\, &=\sum_\lambda\int{d^3k \sum_{i,j}\hat{p}_i^\dagger(\mathbf{k})\hat{p}_j(\mathbf{k})\Big(\lambda(\alpha_i\alpha_j+\tilde\alpha_i\tilde\alpha_j)+i(\beta_i\gamma_j-\tilde\gamma_i\tilde\beta_j+\gamma_i\beta_j-\tilde\beta_i\tilde\gamma_j)\Big)}\\
&+\hat{p}^\dagger_i(\mathbf{k})\hat{p}^\dagger_j(-\mathbf{k})\Big(-\lambda\alpha_i\tilde\alpha_j+i(\gamma_i\tilde\beta_j-\tilde\gamma_i\beta_j)\Big)\\
&+\hat{p}_i(\mathbf{k})\hat{p}_j(-\mathbf{k})\Big(-\lambda\tilde\alpha_i\alpha_j+i(\gamma_i\tilde\beta_j-\tilde\gamma_i\beta_j)\Big),
\end{split}
\end{equation}
where we have made use of the fact that we are free to choose the phases of the coefficients such that $\alpha$ and $\tilde\alpha$ are real. The expressions ~\eqref{eq:algebraicRelations} then immediately imply that $\beta$ and $\tilde\beta$ are purely imaginary, while $\gamma$ and $\tilde\gamma$ are purely real.

We can simplify this further by using Eq.~\eqref{eq:algebraicRelations}  to express $\be_i$, $\bett_i$, $\ga_i$ and $\gatt_i$ in terms of $\al_i$ and $\altt_i$. We may then use the fact that $\altt_{u/d} = 0$ and $\al_p = 0$, as well as the explicit solutions for $\al_{u/d}$ and $\altt_p$ given in Eq.~\eqref{eq:alpm} and Eq.~\eqref{eq:alttp}, to show that the $i\neq j$ terms vanish. 

In order to show this cancellation, we also must make use of the fact that when $i\neq j$, the frequencies $\omega_i$ and $\omega_j$ are related to one another, as both are two distinct roots of the same dispersion relation. The relevant dispersion relations Eqns.~\eqref{eq:dispersionRelationOne}  and ~\eqref{eq:dispersionRelationTwo} are cubics in $\omega$, so sums and products of the two roots which appear can be expressed in terms of the third root. Finally, using Eq.~\eqref{eq:alpm} and Eq.~\eqref{eq:alttp} again, we find Eq.~\eqref{eq:helicityDiagonal} from the main text.

\section{Diagonalisation in non-dual-symmetric media}\label{appendix:generalDiagonalisation}
Here we present explicit expressions for the polariton operators in a non-dual-symmetric medium. It is possible to exactly diagonalise the Hamiltonian (\ref{eq:NonDSHamiltonian}), without making any assumptions about $\omega_E$, $\omega_M$, $\alpha$ or $\beta$. If we do so, we obtain the polariton operators
\begin{equation} \label{eq:achiralPolaritonOperator}
\hat{p}_{\lambda_{i}}(\mathbf{k})=\alpha_i a_\lambda(\mathbf{k})+\tilde\alpha_i a_\lambda^\dagger(-\mathbf{k})+\beta_i b_\lambda(\mathbf{k})+\tilde\beta_i b_\lambda^\dagger(-\mathbf{k})+\gamma_i c_\lambda(\mathbf{k})+\tilde\gamma_i c_\lambda^\dagger(-\mathbf{k}),
\end{equation}
with coefficients
\begin{align} \label{eq:achiralCoeffecient1}
|\alpha_i|^2 &= \Bigg(1-A^2+\frac{\alpha^2 k(1+A)^2\omega_i}{(\omega_i-\tilde\omega_E)^2(\omega_i+\tilde\omega_E)^2}+\frac{\beta^2 k(A-1)^2\omega_i}{(\omega_i-\tilde\omega_M)^2(\omega_i+\tilde\omega_M)^2}\Bigg)^{-1}, \\
\tilde\alpha_i &= A\alpha_i, \\
\beta_i &= \frac{i\alpha}{2(\omega_i-\tilde\omega_E)}\sqrt{\frac{k}{\tilde\omega_E}}(1+A)\alpha_i, \\
\gamma_i &= \frac{\lambda\beta}{2(\omega_i-\tilde\omega_M)}\sqrt{\frac{k}{\tilde\omega_M}}(A-1)\alpha_i, \\
\tilde\beta_i &= \frac{i\alpha}{2(\omega_i+\tilde\omega_E)}\sqrt{\frac{k}{\tilde\omega_E}}(1+A)\alpha_i, \\
\tilde\gamma_i &= \frac{\lambda\beta}{2(\omega_i+\tilde\omega_M)}\sqrt{\frac{k}{\tilde\omega_M}}(A-1)\alpha_i  \label{eq:achiralCoeffecient6}
\end{align}
with
\begin{equation}
A \equiv \Bigg(\frac{\beta^2 k}{2(\omega_i^2-\tilde\omega_M^2)}-\frac{\alpha^2 k}{2(\omega_i^2-\tilde\omega_E^2)}\Bigg) \Bigg(\omega_i+k+\frac{\alpha^2 k}{2(\omega_i^2-\tilde\omega_E^2)}+\frac{\beta^2 k}{2(\omega_i^2-\tilde\omega_M^2)} \Bigg)^{-1}.
\end{equation}
The phases of the coefficients $\alpha_i$ may be chosen freely. In contrast with the dual-symmetric case presented in Appendix \ref{appendix:diagonalisation}, we no longer have that $\alpha_i\neq 0 \implies \tilde{\alpha}_i = 0$, nor do we generally have one negatively refracting branch.
\section{Eigenvectors of the Duality Transform}\label{appendix:dualityTransformEigenvalues}
In this appendix, we discuss the field vectors which are eigenvectors of the duality transformation. We have seen that the helicity per circularly polarised polariton is not $\pm 1$ in non-dual-symmetric media. One way to understand this is by considering, classically, the field vectors which are eigenvectors of the duality transform and the circumstances under which these coincide with the fields of a circularly polarised plane wave in a linear magnetodielectric. 

Firstly, we return to the fact that the duality transform \eqref{eq:dualityTransformAC} can be viewed as a rotation between electric and magnetic quantities. This is most apparent when written in matrix form,
\begin{align}
    \left(\begin{array}{cc}\Ah' \\ \Ch' \end{array}\right)= \left(\begin{array}{cc}
         \cos\theta & \sin\theta \\
         -\sin\theta & \cos\theta 
     \end{array}\right)\left(\begin{array}{cc}\Ah \\ \Ch \end{array}\right), \nonumber\\
    \left(\begin{array}{cc}\Ph' \\ \Mh' \end{array}\right)= \left(\begin{array}{cc}
         \cos\theta & \sin\theta \\
         -\sin\theta & \cos\theta
     \end{array}\right)\left(\begin{array}{cc}\Ph \\ \Mh \end{array}\right).\label{eq:matrixDuality}
\end{align}
The eigenvectors\footnote{There is also a $z$-directional eigenvector of the duality transform, which is not relevant here as we are discussing transverse fields.} of the duality rotation \eqref{eq:matrixDuality} can be written as
\begin{align}\label{eq:dualityVector}
    \mb{e}^\text{duality}_{\la} = \frac{1}{2}\left(\begin{array}{cc}
         \mb{e}_x + i\la\mb{e}_y \\
         - i\la\mb{e}_x + \mb{e}_y
    \end{array}\right)=\frac{1}{2}\left(\begin{array}{cc}
         \mb{e}_\la \\
          -i\la\mb{e}_\la
    \end{array}\right),
\end{align}
with eigenvalues $e^{- i\la\theta}$. We can then connect this to rotations in physical space by considering a circularly polarised plane wave propagating in the $z$-direction. Let us first return to free-space, where an electric field $\mathbf{E}=E_0 \mathbf{e}_\la$ may be described by field vectors $\mathbf{A}_0\propto \mb{e}_\la$ and $\mathbf{C}_0=-i \la\mathbf{A}_0\propto -i \la \mb{e}_\la$, with $\lambda=\pm 1$ referring to the two circular polarisations. The vector $(\Ah_0,\Ch_0)^T$ is then clearly proportional to the duality transform eigenvector \eqref{eq:dualityVector}. These are, as can be seen, closely connected to the eigenvectors of physical rotation about the $z$-direction, which are the circular polarisation vectors $\mb{e}_\pm = \mb{e}_x\pm i\mb{e}_y$, also with eigenvalues $e^{\pm i\theta}$. 

In a dual-symmetric medium, the $\Ah$ and $\Ch$ fields are still related as in free-space and induce polarisations and magnetisations $\mathbf{P}_0\propto \mb{e}_\la$ and $\mathbf{M}_0=-i\la\mathbf{P}_0\propto -i\la\mb{e}_\la$. Because the polarisation and magnetisation have equal amplitude, the vectors $(\Ph_0,\Mh_0)$ of this form are also eigenvectors of the $(\Ph,\Mh)$ transform. In this sense, the helicity quantifies the twist associated with the rotation implied by the circular polarisation vectors.

However, this is not the case in general. In most optical media, the potentials $\Ah$ and $\Ch$ for a circularly polarised wave are no longer related as simply. These then also give rise to polarisation and magnetisation fields that are not related by $\mathbf{M}=-i\la\mathbf{P}$. As an example of this, let us consider a medium with permittivity $\varepsilon$, permeability $\mu$ and susceptibilities $\chi_e$ and $\chi_m$. The potentials are in this case related by $\Ch = -i\la\sqrt{\varepsilon/\mu}\,\Ah$. Also, the polarisation and magnetisation are instead related by $\mathbf{M}=-i\la(\chi_m/\chi_e) \sqrt{\mu/\varepsilon}\, \mathbf{P}$. This does not form an eigenvector of the duality transform, so the circular polarisations are no longer the states of unit helicity. Furthermore, in a dispersive medium where the susceptibilities are functions of frequency, we can see that the scaling between $\mathbf{M}$ and $\mathbf{P}$ will generally be $k$-dependent. This is what causes the helicity of a circularly polarised wave to depend on $k$, as those vectors can always be decomposed in terms of $\mb{e}^\tm{duality}_\pm$ with $k$-dependent amplitudes.\footnote{This is a change of basis, for which the $z$-directional eigenvector is required in the definition. It does not enter in the decomposition, however.}

Another way to put this is that, for a plane wave in a dual-symmetric medium, the rotation \eqref{eq:dualityTransformAC} corresponds to a \textit{physical} rotation of both the fields and the dipoles about the direction of propagation. In a general medium, this is not the case. It is possible to scale the transform as in \eqref{eq:dualityTransformRescaling} to retain the connection between the transformation and rotations about $\mathbf{k}$ in some cases, but this is not possible in general. We choose in this work to use the unscaled transformation in all cases, and to interpret the differences in behaviour between helicity in media and in free space as a manifestation of the fact that the matter and light behave in different ways. We emphasise that, although we have considered linear magnetodielectrics in this discussion, the application of \eqref{eq:dualityTransformAC} and its associated helicity \eqref{eq:helicity} does not depend on any particular constitutive relation connecting the light and matter fields.

\end{document}